\documentclass[twocolumn,aps,prl,epsfig,showpacs,amsmath,amssymb,superscriptaddress]{revtex4}

\usepackage{graphicx}
\usepackage{dcolumn}
\usepackage{bm}

\begin{document}

\title{High Mobility in LaAlO$_{3}$/SrTiO$_{3}$ Heterostructures: Origin, Dimensionality and Perspectives}

\author{G. Herranz}
\affiliation{Unit\'e Mixte de Physique CNRS / Thales, Route
D\'{e}partementale 128, 91767 Palaiseau, France}
\email{gervasi.herranz@thalesgroup.com}
\author{M. Basletic}
\affiliation{Dep. of Physics, Fac. of Science, HR-10002 Zagreb,
Croatia}
\author{M. Bibes}
\affiliation{Institut d'El\'ectronique Fondamentale, Univ.
Paris-Sud, 91405 Orsay, France}
\author{C. Carr\'et\'ero}
\affiliation{Unit\'e Mixte de Physique CNRS / Thales, Route
D\'{e}partementale 128, 91767 Palaiseau, France}
\author{E. Tafra}
\affiliation{Dep. of Physics, Fac. of Science, HR-10002 Zagreb,
Croatia}
\author{E. Jacquet}
\affiliation{Unit\'e Mixte de Physique CNRS / Thales, Route
D\'{e}partementale 128, 91767 Palaiseau, France}
\author{K. Bouzehouane}
\affiliation{Unit\'e Mixte de Physique CNRS / Thales, Route
D\'{e}partementale 128, 91767 Palaiseau, France}
\author{C. Deranlot}
\affiliation{Unit\'e Mixte de Physique CNRS / Thales, Route
D\'{e}partementale 128, 91767 Palaiseau, France}
\author{A. Hamzic}
\affiliation{Dep. of Physics, Fac. of Science, HR-10002 Zagreb,
Croatia}
\author{J.-M. Broto} \affiliation{Laboratoire Nationale
des Champs Magn\'etiques Puls\'es, 143 Av. de Rangueil,31400
Tolouse, France}
\author{A. Barth\'el\'emy}
\affiliation{Unit\'e Mixte de Physique CNRS / Thales, Route
D\'{e}partementale 128, 91767 Palaiseau, France}
\author{A. Fert}
\affiliation{Unit\'e Mixte de Physique CNRS / Thales, Route
D\'{e}partementale 128, 91767 Palaiseau, France}

\date{\today}

\begin{abstract}

\vspace{0.5cm} We have investigated the dimensionality and origin
of the magnetotransport properties of LaAlO$_{3}$ films
epitaxially grown on TiO$_{2}$-terminated SrTiO$_{3}$(001)
substrates. High mobility conduction is observed at low deposition
oxygen pressures ($\rm P_{O2}$ $<10^{-5}$ mbar) and has a
three-dimensional character. However, at higher $\rm P_{O2}$ the
conduction is dramatically suppressed and nonmetallic behavior
appears. Experimental data strongly support an interpretation of
these properties based on the creation of oxygen vacancies in the
SrTiO$_{3}$ substrates during the growth of the LaAlO$_{3}$ layer.
When grown on SrTiO$_{3}$ substrates at low $\rm P_{O2}$, other
oxides generate the same high mobility as LaAlO$_{3}$ films. This
opens interesting prospects for all-oxide electronics.

\vspace{0.5cm}
\end{abstract}

\pacs{73.40.-c, 73.50.Fq, 66.30.Jt}

 \maketitle

Among the different technologies that are currently being
considered to replace CMOS in the coming years, oxide-based
electronics is a promising and rapidly growing field
\cite{ogale05}. Recently, the interest for the electronic
properties of oxide heterointerfaces has been boosted by
experiments on LaAlO$_{3}$/SrTiO$_{3}$ (LAO/STO) samples,
suggesting the existence of a high mobility conducting layer at
the interface between two insulators: a LAO film grown on a STO
substrate at pressures smaller than 10$^{-4}$ mbar \cite{ohtomo04,
hwang04, siemons06, huijben06, thiel06, schneider06}.

For such structures, two types of interface can be defined
depending on the termination (AO or BO$_{2}$) of the perovskite
ABO$_{3}$ blocks. The LaO/TiO$_{2}$ interface was found to be
conductive (n-type) with a large low temperature Hall mobility (up
to $\mu_{H} \geq 10^{4}$ cm$^{2}$/Vs) and a large sheet resistance
ratio (R$\rm_{sheet}(300 K)$/R$\rm_{sheet}(5 K) \sim 10^{3}$)
\cite{ohtomo04, hwang04, siemons06, huijben06, thiel06}. On the
other hand, the AlO$_{2}$/SrO interface was found insulating. In
LAO the La and Al cations have both a 3+ valence, and in STO Sr is
2+ and Ti is 4+. It is argued that the resulting interface
electronic discontinuity is avoided through charge transfer from
LAO to STO \cite{nakagawa06}. Within this picture, the high
mobility gas is formed when the LAO layer provides the
LaO/TiO$_{2}$ interface with half an electron per two-dimensional
unit cell, corresponding to n$\rm_{sheet} \approx
6.25\times10^{14}$ cm$^{-2}$. According to this model, the
electronic properties should not depend on the oxygen pressure
used to grow the LAO film on the TiO$_{2}$-terminated STO
substrate. However, up to now, all the reported high-mobility
LAO/STO structures have been grown on STO substrates in reducing
conditions (P$\rm_{O2} \leq 10^{-4}$ mbar) \cite{ohtomo04,
hwang04, siemons06, huijben06, thiel06}.

In parallel, it has recently been demonstrated that the growth of
other oxides like Co-doped (La, Sr)TiO$_{3}$ (CoLSTO)
\cite{herranz06} or the homoepitaxial growth of SrTiO$_{3}$ films
on STO substrates at low pressure (P$\rm_{O2} < 10^{-4}$ mbar)
\cite{ohtomo06} also generates a high mobility conduction, with
the same transport properties as the above-discussed LAO/STO
structures. However, in those cases charge transfer effects are
either different from the LAO/STO case or formally absent. Indeed,
the high-mobility conduction in CoLSTO/STO films has been
demonstrated to be due to oxygen vacancy-doping of the STO
substrates \cite{herranz06}. Therefore, in order to test whether
charge transfer across the interface is really responsible for the
high-mobility conductive behavior, one should fabricate LAO/STO
structures in conditions that prevent the doping of STO substrates
by oxygen vacancies, i.e., at pressures well above 10$^{-4}$ mbar
\cite{herranz06} and compare these properties with those of
structures prepared at smaller pressure.

To address this issue, we have measured the magnetotransport
properties of LAO (6-20) nm-thick films grown by pulsed laser
deposition in a wide range of pressures (10$^{-6}$ - 10$^{-3}$
mbar) on 0.5-1 mm-thick TiO$_{2}$-terminated STO(001)
single-crystal substrates (for details about magnetotransport
experiments see Ref. \cite{herranz06}). The interface was
contacted with Al/Au through the LAO layer by locally etching the
LAO with accelerated Ar ions down to the interface, in a chamber
equipped with a secondary ion mass spectroscopy detection system.
The interface has been characterized by aberration-corrected
high-resolution transmission electron microscopy (HRTEM), and
found that the LAO layer is fully strained, and the interface is
close to atomically sharp, with no dislocations (the details of
HRTEM characterization are reported in Ref. \cite{maurice06b}). We
note that films cooled down in a high pressure P$\rm_{O2} \approx
300$ mbar from the deposition (T = 750 $^\circ$C) to room
temperature were all insulating or highly resistive ($>100$
M$\Omega$ at room temperature). Consequently, in the following we
will report on films cooled down to room temperature in the
deposition pressure.

\begin{figure}[!h]
\vspace{0.1cm}
 \includegraphics[keepaspectratio=true,width=\columnwidth]{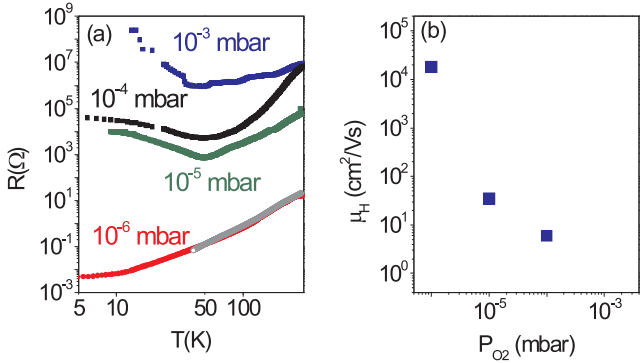}
 \caption{(a) T-dependence of the resistance of samples grown
 at $\rm P_{O2}$ = 10$^{-6}$ - 10$^{-3}$ mbar;
 grey symbols correspond to the sample grown at 10$^{-6}$ mbar
 after removing the LAO film by mechanical polishing (see text)
 and (b) dependence of the mobility at 4K $\mu_{H,4K}$ on the deposition pressure.}
 \label{pressure}
\end{figure}

In Fig. \ref{pressure} we present the dependence on P$\rm_{O2}$ of
the transport properties of LAO/STO samples with thickness
\textit{t} = 20 nm. The temperature (T) dependence of the
resistance and mobility of our LAO/STO samples (Fig.
\ref{pressure}a) grown at low pressure (P$\rm_{O2} < 10^{-5}$
mbar) are similar to those reported in other works \cite{ohtomo04,
hwang04, siemons06, huijben06, thiel06}. We observe that the
samples become less conductive as P$\rm_{O2}$ is higher (Fig.
\ref{pressure}a). For films grown at P$\rm_{O2} \geq 10^{-5}$ mbar
we observe the presence of resistance upturns with nonmetallic
behavior below those temperatures. Additionally, the mobility at 4
K ($\mu_{H,4K}$) decreases drastically as P$\rm_{O2}$ is increased
(Fig. \ref{pressure}b), and samples grown at P$\rm_{O2} \geq
10^{-4}$ mbar show $\mu_{H,4K} < 10$ cm$^{2}$/Vs (for sample grown
at 10$^{-3}$ mbar the mobility is not measurable). Both
observations are at odds with an interpretation based on charge
transfer effects.

\begin{figure}[!h]
\vspace{0.1cm}
 \includegraphics[keepaspectratio=true,width=\columnwidth]{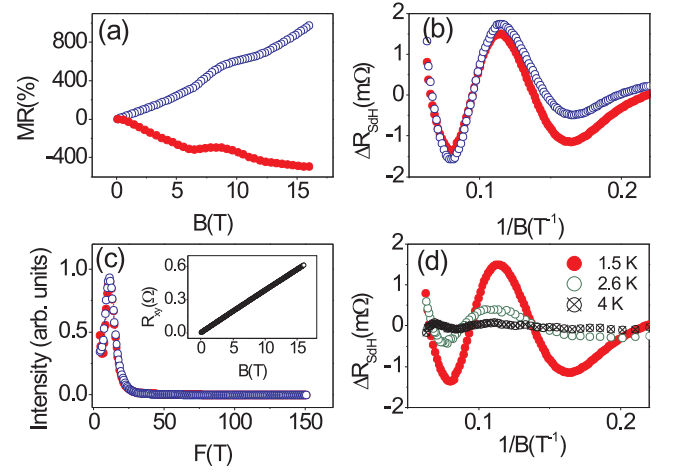}
 \caption{(a) Magnetic field dependence of the perpendicular (open symbols)
 and longitudinal (closed symbols) magnetoresistances at 1.5 K.
 (b) Oscillatory part of the magnetoresistance  $\Delta$R$\rm_{SdH}$.
 (c) Spectral power density from FFT analysis of
 $\Delta$R$\rm_{SdH}$(H). (d) $\Delta$R$\rm_{SdH}$(H) corresponding to perpendicular
 magnetoresistance at different T.
 The inset of (c) shows the field dependence of the
 Hall resistance R$_{xy}$ at 4.2 K.}
 \label{sdh}
\end{figure}

We proceed now to the analysis of the dimensionality and
determination of the thickness of the metallic gas in our LAO/STO
sample (\textit{t} = 20 nm) grown at P$\rm_{O2}$ = 10$^{-6}$ mbar.
For this sample $\mu_{H,4K}$ = 1.8$\times$10$^{4}$ cm$^{2}$/Vs and
R$\rm_{sheet, 4K}$ $\sim 10^{-2} \Omega/\square$, in good
agreement with data reported in Ref. \cite{ohtomo04}.
Magnetoresistance curves, MR =
[R$_{xx}$(H)-R$_{xx}$(0)]/R$_{xx}$(0), were measured with the
field H either perpendicular to the sample plane (PMR), or
parallel to the current in the sample plane (LMR). Both
magnetoresistances exhibit Shubnikov - de Haas (SdH) oscillations
at T $<$ 4 K and magnetic fields B $\geq$ 6 T (the data at 1. 5 K
are shown in Fig. \ref{sdh}a). Note that these oscillations
disappear progressively at T $>$ 1.5 K (see Fig. \ref{sdh}d), as
expected. After subtracting the background contribution, the SdH
oscillations are even more clearly seen in Fig. \ref{sdh}b. From
this figure, it is evident that the period of the oscillations
does not depend on the field orientation. The observation of
similar SdH oscillations in both configurations excludes any
interfacial confinement of carriers.

The SdH oscillations were analyzed following the protocol
described in detail in Ref. \cite{herranz06}. The Fast Fourier
Transform procedure confirmed that the SdH frequency F$\rm_{SdH}$
(which is related to the cross-sectional area $A_{ext}$ of
extremal electronic orbits in k-space perpendicular to the applied
field by F$\rm_{SdH}$=$\hbar$A$_{ext}$/2e) is the same for the PMR
and LMR configurations - cf. Fig. \ref{sdh}c. This definitely
demonstrates that the electronic system is 3D and homogeneous.  A
system with a non-uniform carrier density (varying as a function
of the distance from the film/substrate interface) would lead, in
the PMR configuration, to a superposition of different frequencies
and to a broadening or blurring of the spectrum.

Furthermore, our data enable us to estimate the thickness of the
high mobility gas. The simplest approximation is to suppose a
spherical Fermi surface; in this case, from F$\rm_{SdH}$ $\approx$
11.1 T (cf. Fig. \ref{sdh}c), we obtain k$\rm_{F}$ $\approx$
1.8$\times$10$^{6}$ cm$^{-1}$. Since the electronic system is 3D,
its thickness is thus much larger than the Fermi wavelength
$\lambda\rm_{F}$ = 2$\pi$/k$\rm_{F}$ $\approx$ 35 nm. To determine
this thickness more precisely, we have used our Hall resistance
$R_{xy}$ (B) data (inset of Fig. \ref{sdh}c). Defining the sheet
carrier density $n_{sheet}$ as the product of the carrier density
$n$ and the thickness of the metallic region $t_{hm}$, we get
$n_{sheet} = n \times t_{hm}= B/(e·R_{xy}) \approx$
1.6$\times10^{16}$ cm$^{-2}$. The carrier density $n$ can be
determined independently by using the value of k$\rm_{F}$ (derived
from F$\rm_{SdH}$); with $n$ = k$\rm _{F}^{3}/3\pi^{2}$ $\approx$
2$\times$10$^{17}$ cm$^{-3}$, one finally obtains $t_{hm} \approx$
800 $\mu$m. Now, if instead of an ideal spherical Fermi surface we
use the more realistic k-space geometry of doped STO (as described
in the Appendix of Ref. \cite{herranz06}), we find k$\rm_{F}$
$\approx$ 0.9$\times$10$^{6}$ cm$^{-1}$, $\lambda_{F}$ $\approx$
70 nm, $n \approx$ $3\times$10$^{17}$ cm$^{-3}$ and finally
$t_{hm} \approx$ 530 $\mu$m which is strikingly close to the
substrate thickness. Our analysis unambiguously proves that the
high-mobility transport properties of the LAO/STO sample grown at
P$\rm_{O2}$ = 10$^{-6}$ mbar are due to a conducting region
homogeneously extending over hundreds of $\mu$m inside the STO
substrate. This conclusion is further confirmed by the observation
of the same T-dependence (see Fig. \ref{pressure}a) of the
resistance for a 10$^{-6}$ mbar sample before and after removing
the film and about 15 $\mu$m of the STO substrate by mechanical
polishing on the film side (Fig. \ref{pressure}a). This is
additional evidence that the transport properties of LAO/STO
samples are in fact due to the STO substrate.

\begin{figure}[!h]
\vspace{0.1cm}
 \includegraphics[keepaspectratio=true,width=\columnwidth]{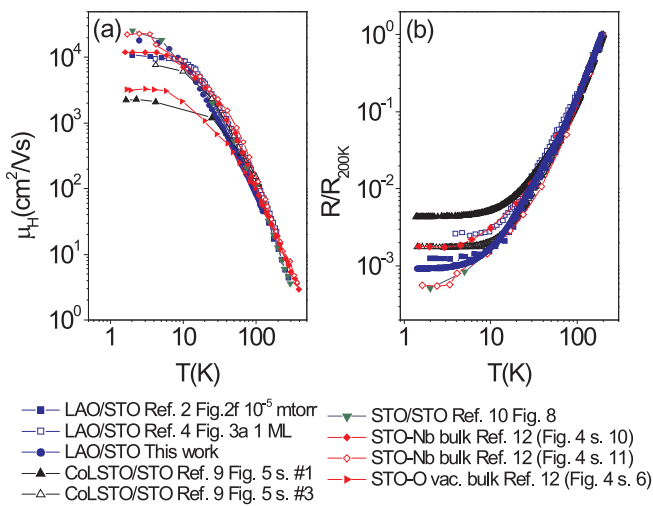}
 \caption{(a) T-dependence of the Hall mobility $\mu_{H}$ and
 (b) T-dependence of the resistance normalized to the
 value at 200 K; both for CoLSTO/STO, LAO/STO, STO$_{3-\delta}$/STO samples
 and doped STO single crystals.}
 \label{sro}
\end{figure}

For films grown at P$\rm_{O2}$ $\geq$ 10$^{-5}$ mbar, the sheet
resistance increases (R$\rm_{sheet} >$ 1 k$\Omega$  even at low
temperature), and consequently SdH oscillations cannot be
observed. However, some insight can be gained by comparing the
transport data of LAO/STO structures to those reported for doped
bulk STO single crystals. We have collected in Fig. \ref{sro} the
T-dependence of the mobility (Fig. \ref{sro}a) and normalized
resistance (Fig. \ref{sro}b) for STO bulk single crystals (doped
or treated in reducing atmospheres) \cite{tufte67}, LAO films
grown on STO substrates at P$\rm_{O2} <$ 10$^{-4}$ mbar (this work
and Refs. \cite{ohtomo04} and \cite{siemons06}), Co-doped
(La,Sr)TiO$_{3}$ (Co-LSTO) films grown at low P$\rm_{O2}$ on STO
substrates (Ref. \cite{herranz06}) and STO$\rm_{3-\delta}$/STO
homoepitaxial films \cite{ohtomo06}. Let us point out here that
one can extract $\mu_{H}$ from the Hall experiments without
knowing the thickness of the metallic system. This allows a
reliable comparison of data from different sources. The main
characteristic of both T-dependences (mobility - Fig. \ref{sro}a
and normalized resistance - Fig. \ref{sro}b) is the remarkable
resemblance of the behavior for the LAO/STO samples and bulk STO
specimens doped with Nb-, La- or oxygen vacancies \cite{tufte67,
frederikse67, lee71, frederikse64, olaya02}.

\begin{figure}[!h]
\vspace{0.1cm}
 \includegraphics[keepaspectratio=true,width=\columnwidth]{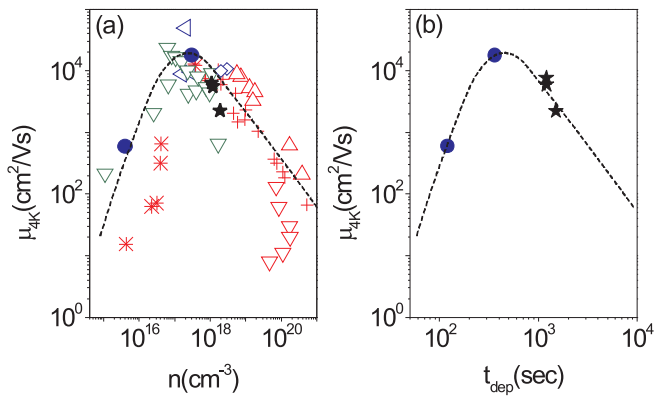}
 \caption{(a) Dependence of $\mu_{H,4K}$ on the carrier density:
 STO single crystals doped with oxygen
 vacancies (+) or Nb ($\bigtriangleup$) [Ref. \cite{frederikse67}];
 slightly reduced STO single crystals ($\ast$) [Ref. \cite{lee71}];
 La-doped STO thin films from ($\bigtriangledown$)
 [Ref. \cite{olaya02}];
 CoLSTO/STO samples grown at low pressure [Ref. \cite{herranz06}
 ($\star$)];
 STO$_{3-\delta}$/STO films grown at low pressure
 ($\bigtriangledown$)
 [Ref. \cite{ohtomo06}];
 LAO/STO samples grown at low pressure [Ref. \cite{siemons06}]
 ($\triangleleft$),
 [Ref. \cite{ohtomo04}]
 ($\diamond$)
 and  this work
 ($\bullet$).
 The dashed lines are guides for the eye.
 (b) Dependence on time deposition t$_{dep}$ of $\mu_{4K}$
 in Co-LSTO (Ref. \cite{herranz06}) and LAO films (this work)
 grown on STO substrates at low $\rm P_{O2}$.}
 \label{tdep}
\end{figure}

The dependence of the mobility $\mu_{H}$ on the carrier density
$n$ for different doped STO single crystals \cite{frederikse67,
lee71} is presented in Fig. \ref{tdep}a. The highest mobilities
are observed when the carrier density is $\sim$ 3$\times$10$^{17}$
cm$^{-3}$. The mobility decays rapidly as the $n$ increases; this
is due to the introduction of extrinsic impurities associated to
potential fluctuations induced in the lattice. For concentrations
below 10$^{17}$ cm$^{-3}$, the impurity band formed by the donors
has a reduced width (due to a decrease of the overlap of the
impurity wave functions), and the mobility decreases as well. Fig.
\ref{tdep}a also includes our data for Co-LSTO/STO (Ref.
\cite{herranz06}) and LAO/STO samples which perfectly match the
bulk STO data. The figure is completed by the additional
recompilation of the data for STO$\rm_{3-\delta}$/STO
homoepitaxial films \cite{ohtomo06} and other LAO/STO samples
(Refs. \cite{ohtomo04} and \cite{siemons06}), considering that
carriers extend over a region of 500 $\mu$m. The similarity
between the transport properties of LAO/STO samples exhibiting
high mobility (grown at P$\rm_{O2}$ $< 10^{-4}$ mbar) and those of
reduced STO bulk single crystals, suggests that the high mobility
of LAO/STO samples might arise from the doping of STO with oxygen
vacancies. This conclusion is also supported by recent cathodo-
and photoluminescence studies of Kalabukhov et al.
\cite{kalabukhov06} and by ultraviolet photoelectron spectroscopy
of Siemons et al. \cite{siemons06}.

A remaining question is how far the oxygen vacancies can diffuse
into the STO substrates during film deposition. It is known that
the exchange of oxygen is strongly enhanced at the film/substrate
interface as compared to the substrate/vacuum interface
\cite{leonhardt02}. Their diffusion coefficient $D_{V}$ has been
measured with different methods \cite{ishigaki88} and at various
P$\rm_{O2}$ and T. For T $\approx$ 750 $^\circ$C and at pressures
as high as P$\rm_{O2}$ $\approx$ 10$^{-2}$ mbar, values of
10$^{-4}$ - 10$^{-5}$ cm$^{2}$/s were found. During a time
interval \textit{t} the vacancies diffuse along a distance
l$_{Ovac} \approx$ $(D_{V}t)^{1/2}$, which is, for $t$ = 10 s
(i.e. typical deposition times for the thinnest films) between 100
$\mu$m -300 $\mu$m. It is then reasonable to assume that oxygen
vacancies (created when STO is exposed to reducing atmospheres
during the film growth) can extend to a region within the STO
substrate compatible with that found from the SdH analysis.

From these considerations, it should be expected that the
transport properties are modulated by the effective deposition
time $t_{dep}$ and, in turn, by film thickness. Fig. \ref{tdep}b
shows the dependence on $t_{dep}$ of $\mu_{H,4K}$ of CoLSTO and
LAO films deposited at P$\rm_{O2} \approx 10^{-6}$ mbar. We
observe that this $\mu_{H,4K}$ - $t_{dep}$ curve bears a striking
resemblance with the $\mu_{H}$ - $n$ curve for the same samples
plotted in Fig. \ref{tdep}a. This is a strong indication that
$t_{dep}$ increases the density of carriers in the STO substrates.
This should also apply at slightly higher pressures P$\rm_{O2}
\approx 10^{-5}$ mbar, used for growing LAO/STO structures which
were reported to have modulated transport properties upon
increasing the deposited film thickness \cite{huijben06, thiel06}.
Indeed, in an oxygen vacancy-doping scenario, an insulator-metal
transition (IMT) should be expected as a function of $t_{dep}$
and, therefore, of the number of LAO unit cells grown on the STO
substrate \cite{huijben06, thiel06}. When the deposited film
thickness increases, the critical concentration $n_{c}$ for
impurity band formation is reached and an IMT occurs. This
behavior is well known in group IV semiconductors doped with
impurities \cite{alexander68}. This IMT is accompanied by an
abrupt conductivity jump that can be estimated as
$\Delta\sigma_{c} = n_{c}\times e\times\mu_{c}$, with $n_{c}
\approx 10^{15}$ cm$^{-3}$ for doped STO (Fig. \ref{tdep}a) and
$\mu_{c}\approx 1$ cm$^{2}/Vs$ is the mobility at this critical
concentration \cite{lee71}. Substituting values we obtain
$\Delta\sigma_{c} \approx 1.6\times10^{-4} \Omega^{-1}$cm$^{-1}$.
If one assumes a conducting thickness $l \approx 100 \mu$m, one
should observe a conductance jump of $\Delta G_{c} \approx
\Delta\sigma_{c}\times l \approx 10^{-5} \Omega^{-1}$ across the
IMT. A similar value of $\Delta G_{c}$ was reported for the IMT at
a critical thickness of LAO films deposited on STO substrates
\cite{thiel06}.

In summary, we have brought proof of an alternative and coherent
explanation for the high-mobility in LAO/STO structures based on
the doping of STO substrates with oxygen vacancies. This doping is
achieved efficiently by the growth of LAO films on STO substrates.
This remarkable result might be used to fabricate heterostructures
integrating multifunctional oxides with high mobility STO. E.g.,
spin injection into a STO channel might be achieved with
spin-polarized sources such as CoLSTO \cite{herranz06,
herranz06prl}. Also, an interesting possibility is to dope STO
below the critical concentration $n_{c}$, and to activate the
conduction through application of electrical fields. Recent
reports on STO channel field-effect transistors with organic
\cite{nakamura06} or LAO \cite{thiel06} gate insulators evidence
enormous field-effects in such structures. All these possibilities
open interesting prospects and should stimulate further studies of
all-oxide structures for spintronics.

\acknowledgements{G. Herranz acknowledges financial support from
the DURSI (Generalitat de Catalunya, Spain). Financial supports
from PAI-France-Croatia COGITO Program No. 82/240083, and MoSES
project No. 119-1191458-1023, are acknowledged. We acknowledge
experimental support from Y. Lema\^itre, S. Fusil and B. Raquet.}

\vspace{0.5em}

\bibliography{biblio2}

\begin{thebibliography}{22}
\expandafter\ifx\csname natexlab\endcsname\relax\def\natexlab#1{#1}\fi
\expandafter\ifx\csname bibnamefont\endcsname\relax
  \def\bibnamefont#1{#1}\fi
\expandafter\ifx\csname bibfnamefont\endcsname\relax
  \def\bibfnamefont#1{#1}\fi
\expandafter\ifx\csname citenamefont\endcsname\relax
  \def\citenamefont#1{#1}\fi
\expandafter\ifx\csname url\endcsname\relax
  \def\url#1{\texttt{#1}}\fi
\expandafter\ifx\csname urlprefix\endcsname\relax\def\urlprefix{URL }\fi
\providecommand{\bibinfo}[2]{#2}
\providecommand{\eprint}[2][]{\url{#2}}

\bibitem[{\citenamefont{Ogale}(2005)}]{ogale05}
\bibinfo{author}{\bibfnamefont{S.~B.} \bibnamefont{Ogale}},
  \emph{\bibinfo{title}{Thin Films and Heterostructures for Oxide Electronics}}
  (\bibinfo{publisher}{Springer Verlag}, \bibinfo{year}{2005}).

\bibitem[{\citenamefont{Ohtomo and Hwang}(2004)}]{ohtomo04}
\bibinfo{author}{\bibfnamefont{A.}~\bibnamefont{Ohtomo}} \bibnamefont{and}
  \bibinfo{author}{\bibfnamefont{H.~Y.} \bibnamefont{Hwang}},
  \bibinfo{journal}{Nature} \textbf{\bibinfo{volume}{427}},
  \bibinfo{pages}{423} (\bibinfo{year}{2004}).

\bibitem[{\citenamefont{H.Y.Hwang et~al.}(2004)\citenamefont{H.Y.Hwang, Ohtomo,
  Nakagawa, Muller, and Grazul}}]{hwang04}
\bibinfo{author}{\bibnamefont{H.Y.Hwang}},
  \bibinfo{author}{\bibfnamefont{A.}~\bibnamefont{Ohtomo}},
  \bibinfo{author}{\bibfnamefont{N.}~\bibnamefont{Nakagawa}},
  \bibinfo{author}{\bibfnamefont{D.}~\bibnamefont{Muller}}, \bibnamefont{and}
  \bibinfo{author}{\bibfnamefont{J.}~\bibnamefont{Grazul}},
  \bibinfo{journal}{Physica E} \textbf{\bibinfo{volume}{22}},
  \bibinfo{pages}{712} (\bibinfo{year}{2004}).

\bibitem[{\citenamefont{Siemons et~al.}(2006)}]{siemons06}
\bibinfo{author}{\bibfnamefont{W.}~\bibnamefont{Siemons}} \bibnamefont{et~al.},
  \bibinfo{journal}{cond-mat/0603598}  (\bibinfo{year}{2006}).

\bibitem[{\citenamefont{Huijben et~al.}(2006)}]{huijben06}
\bibinfo{author}{\bibfnamefont{M.}~\bibnamefont{Huijben}} \bibnamefont{et~al.},
  \bibinfo{journal}{Nature Materials} \textbf{\bibinfo{volume}{5}},
  \bibinfo{pages}{556} (\bibinfo{year}{2006}).

\bibitem[{\citenamefont{Thiel et~al.}(2006)\citenamefont{Thiel, Hammerl,
  Schmehl, Schneider, and Mannhart}}]{thiel06}
\bibinfo{author}{\bibfnamefont{S.}~\bibnamefont{Thiel}},
  \bibinfo{author}{\bibfnamefont{G.}~\bibnamefont{Hammerl}},
  \bibinfo{author}{\bibfnamefont{A.}~\bibnamefont{Schmehl}},
  \bibinfo{author}{\bibfnamefont{C.~W.} \bibnamefont{Schneider}},
  \bibnamefont{and} \bibinfo{author}{\bibfnamefont{J.}~\bibnamefont{Mannhart}},
  \bibinfo{journal}{Science} \textbf{\bibinfo{volume}{313}},
  \bibinfo{pages}{1942} (\bibinfo{year}{2006}).

\bibitem[{\citenamefont{Schneider et~al.}(2006)\citenamefont{Schneider, Thiel,
  Hammerl, Richter, and Mannhart}}]{schneider06}
\bibinfo{author}{\bibfnamefont{C.~W.} \bibnamefont{Schneider}},
  \bibinfo{author}{\bibfnamefont{S.}~\bibnamefont{Thiel}},
  \bibinfo{author}{\bibfnamefont{G.}~\bibnamefont{Hammerl}},
  \bibinfo{author}{\bibfnamefont{C.}~\bibnamefont{Richter}}, \bibnamefont{and}
  \bibinfo{author}{\bibfnamefont{J.}~\bibnamefont{Mannhart}},
  \bibinfo{journal}{Appl. Phys. Lett.} \textbf{\bibinfo{volume}{89}},
  \bibinfo{pages}{122101} (\bibinfo{year}{2006}).

\bibitem[{\citenamefont{Nakagawa et~al.}(2006)\citenamefont{Nakagawa, Hwang,
  and Muller}}]{nakagawa06}
\bibinfo{author}{\bibfnamefont{N.}~\bibnamefont{Nakagawa}},
  \bibinfo{author}{\bibfnamefont{H.~Y.} \bibnamefont{Hwang}}, \bibnamefont{and}
  \bibinfo{author}{\bibfnamefont{D.~A.} \bibnamefont{Muller}},
  \bibinfo{journal}{Nature Materials} \textbf{\bibinfo{volume}{5}},
  \bibinfo{pages}{204} (\bibinfo{year}{2006}).

\bibitem[{\citenamefont{Herranz et~al.}(2006{\natexlab{a}})}]{herranz06}
\bibinfo{author}{\bibfnamefont{G.}~\bibnamefont{Herranz}} \bibnamefont{et~al.},
  \bibinfo{journal}{Phys. Rev. B} \textbf{\bibinfo{volume}{73}},
  \bibinfo{pages}{064403} (\bibinfo{year}{2006}{\natexlab{a}}).

\bibitem[{\citenamefont{Ohtomo and Hwang}(2006)}]{ohtomo06}
\bibinfo{author}{\bibfnamefont{A.}~\bibnamefont{Ohtomo}} \bibnamefont{and}
  \bibinfo{author}{\bibfnamefont{H.}~\bibnamefont{Hwang}},
  \bibinfo{journal}{cond-mat/0604117}  (\bibinfo{year}{2006}).

\bibitem[{\citenamefont{Maurice et~al.}(2006)}]{maurice06b}
\bibinfo{author}{\bibfnamefont{J.-L.} \bibnamefont{Maurice}}
  \bibnamefont{et~al.}, \bibinfo{journal}{Phys. Stat. Sol.}
  \textbf{\bibinfo{volume}{203}}, \bibinfo{pages}{2145} (\bibinfo{year}{2006}).

\bibitem[{\citenamefont{Tufte and Chapman}(1967)}]{tufte67}
\bibinfo{author}{\bibfnamefont{O.~N.} \bibnamefont{Tufte}} \bibnamefont{and}
  \bibinfo{author}{\bibfnamefont{P.~W.} \bibnamefont{Chapman}},
  \bibinfo{journal}{Phys. Rev.} \textbf{\bibinfo{volume}{155}},
  \bibinfo{pages}{796} (\bibinfo{year}{1967}).

\bibitem[{\citenamefont{Frederikse and Hosler}(1967)}]{frederikse67}
\bibinfo{author}{\bibfnamefont{H.~P.~R.} \bibnamefont{Frederikse}}
  \bibnamefont{and} \bibinfo{author}{\bibfnamefont{W.~R.}
  \bibnamefont{Hosler}}, \bibinfo{journal}{Phys. Rev.}
  \textbf{\bibinfo{volume}{161}}, \bibinfo{pages}{822} (\bibinfo{year}{1967}).

\bibitem[{\citenamefont{Lee et~al.}(1971)\citenamefont{Lee, Yahia, and
  Brebner}}]{lee71}
\bibinfo{author}{\bibfnamefont{C.}~\bibnamefont{Lee}},
  \bibinfo{author}{\bibfnamefont{J.}~\bibnamefont{Yahia}}, \bibnamefont{and}
  \bibinfo{author}{\bibfnamefont{J.~L.} \bibnamefont{Brebner}},
  \bibinfo{journal}{Phys. Rev. B} \textbf{\bibinfo{volume}{3}},
  \bibinfo{pages}{2525} (\bibinfo{year}{1971}).

\bibitem[{\citenamefont{Frederikse et~al.}(1964)\citenamefont{Frederikse,
  Thurber, and Hosler}}]{frederikse64}
\bibinfo{author}{\bibfnamefont{H.~P.~R.} \bibnamefont{Frederikse}},
  \bibinfo{author}{\bibfnamefont{W.~R.} \bibnamefont{Thurber}},
  \bibnamefont{and} \bibinfo{author}{\bibfnamefont{W.~R.}
  \bibnamefont{Hosler}}, \bibinfo{journal}{Phys. Rev.}
  \textbf{\bibinfo{volume}{134}}, \bibinfo{pages}{A442} (\bibinfo{year}{1964}).

\bibitem[{\citenamefont{Olaya et~al.}(2002)\citenamefont{Olaya, Pan, Rogers,
  and Price}}]{olaya02}
\bibinfo{author}{\bibfnamefont{D.}~\bibnamefont{Olaya}},
  \bibinfo{author}{\bibfnamefont{F.}~\bibnamefont{Pan}},
  \bibinfo{author}{\bibfnamefont{C.~T.} \bibnamefont{Rogers}},
  \bibnamefont{and} \bibinfo{author}{\bibfnamefont{J.~C.} \bibnamefont{Price}},
  \bibinfo{journal}{Appl. Phys. Lett.} \textbf{\bibinfo{volume}{80}},
  \bibinfo{pages}{2928} (\bibinfo{year}{2002}).

\bibitem[{\citenamefont{Kalabukhov et~al.}(2007)}]{kalabukhov06}
\bibinfo{author}{\bibfnamefont{A.~S.} \bibnamefont{Kalabukhov}}
  \bibnamefont{et~al.}, \bibinfo{journal}{Phys. Rev. B}
  \textbf{\bibinfo{volume}{75}}, \bibinfo{pages}{121404(R)}
  (\bibinfo{year}{2007}).

\bibitem[{\citenamefont{Leonhardt et~al.}(2002)\citenamefont{Leonhardt, Souza,
  Claus, and Maier}}]{leonhardt02}
\bibinfo{author}{\bibfnamefont{M.}~\bibnamefont{Leonhardt}},
  \bibinfo{author}{\bibfnamefont{R.~D.} \bibnamefont{Souza}},
  \bibinfo{author}{\bibfnamefont{J.}~\bibnamefont{Claus}}, \bibnamefont{and}
  \bibinfo{author}{\bibfnamefont{J.}~\bibnamefont{Maier}}, \bibinfo{journal}{J.
  Electrochem Soc.} \textbf{\bibinfo{volume}{149}}, \bibinfo{pages}{J19}
  (\bibinfo{year}{2002}).

\bibitem[{\citenamefont{Ishigaki et~al.}(1988)\citenamefont{Ishigaki, Yamauchi,
  Kishio, Mizusaki, and Fueki}}]{ishigaki88}
\bibinfo{author}{\bibfnamefont{T.}~\bibnamefont{Ishigaki}},
  \bibinfo{author}{\bibfnamefont{S.}~\bibnamefont{Yamauchi}},
  \bibinfo{author}{\bibfnamefont{K.}~\bibnamefont{Kishio}},
  \bibinfo{author}{\bibfnamefont{J.}~\bibnamefont{Mizusaki}}, \bibnamefont{and}
  \bibinfo{author}{\bibfnamefont{K.}~\bibnamefont{Fueki}},
  \bibinfo{journal}{J.Solid State Chem.} \textbf{\bibinfo{volume}{73}},
  \bibinfo{pages}{179} (\bibinfo{year}{1988}).

\bibitem[{\citenamefont{Alexander and Holcomb}(1968)}]{alexander68}
\bibinfo{author}{\bibfnamefont{M.~N.} \bibnamefont{Alexander}}
  \bibnamefont{and} \bibinfo{author}{\bibfnamefont{D.~F.}
  \bibnamefont{Holcomb}}, \bibinfo{journal}{Rev. of Mod. Phys.}
  \textbf{\bibinfo{volume}{40}}, \bibinfo{pages}{815} (\bibinfo{year}{1968}).

\bibitem[{\citenamefont{Herranz et~al.}(2006{\natexlab{b}})}]{herranz06prl}
\bibinfo{author}{\bibfnamefont{G.}~\bibnamefont{Herranz}} \bibnamefont{et~al.},
  \bibinfo{journal}{Phys. Rev. Lett.} \textbf{\bibinfo{volume}{96}},
  \bibinfo{pages}{027207} (\bibinfo{year}{2006}{\natexlab{b}}).

\bibitem[{\citenamefont{Nakamura et~al.}(2006)}]{nakamura06}
\bibinfo{author}{\bibfnamefont{H.}~\bibnamefont{Nakamura}}
  \bibnamefont{et~al.}, \bibinfo{journal}{Appl. Phys. Lett.}
  \textbf{\bibinfo{volume}{89}}, \bibinfo{pages}{133504}
  (\bibinfo{year}{2006}).

\end{thebibliography}

\end{document}